# NbIr$_2$B$_2$ and TaIr$_2$B$_2$ – new low symmetry noncentrosymmetric superconductors with strong spin orbit coupling

Karolina Górnicka[1,2], Xin Gui[3], Bartlomiej Wiendlocha[4], Loi T. Nguyen[3],

Weiwei Xie[5], Robert J. Cava[3] and Tomasz Klimczuk[1,2,*]

[1] Faculty of Applied Physics and Mathematics, Gdansk University of Technology,

ul. Narutowicza 11/12, 80-233 Gdańsk, Poland, e-mail: karolina.gornicka@pg.edu.pl

[2] Advanced Materials Centre, Gdansk University of Technology,

ul. Narutowicza 11/12, 80-233 Gdańsk, Poland,

[3] Department of Chemistry, Princeton University, New Jersey NJ 08540

[4] Faculty of Physics and Applied Computer Science, AGH University of Science and Technology,

Aleja Mickiewicza 30, 30-059 Kraków, Poland

[5] Department of Chemistry and Chemical Biology, Rutgers University, Piscataway, NJ 08854, USA

* Tomasz Klimczuk, e-mail: tomasz.klimczuk@pg.edu.pl

**Superconductivity was first observed more than a century ago, but the search for new superconducting materials remains a challenge. The Cooper pairs in superconductors are ideal embodiments of quantum entanglement. Thus, novel superconductors can be critical for both learning about electronic systems in condensed matter and for possible application in future quantum technologies. Here we present two previously unreported materials, NbIr$_2$B$_2$ and TaIr$_2$B$_2$, with superconducting transitions at 7.2 K and 5.2 K, respectively. They display a unique noncentrosymmetric crystal structure, and for both compounds the magnetic field that**

**destroys the superconductivity at 0 K exceeds one of the fundamental characteristics of conventional superconductors (the "Pauli limit"), suggesting that the superconductivity may be unconventional. Supporting this experimentally based deduction, first-principle calculations show a spin-split Fermi surface due to the presence of strong spin-orbit coupling. These materials may thus provide an excellent platform for the study of non-BCS superconductivity in intermetallic compounds.**

1. Introduction

Superconducting compounds continue to challenge our ideas about how to understand the behavior of electronic materials. Across the many superconductors known, there are two fundamental parameters of most general interest: the temperature below which the superconducting state occurs ($T_c$) and the critical magnetic field required to fully suppress the superconductivity ($H_{c2}$). The second parameter, which is crucial from the practical point of view, must, at 0 Kelvin in a conventional picture, be below the so called "Pauli limit" ($\mu_0 H_{c2}(0) = 1.85 T_c$), which is derived from a simple relation assuming that the Zeeman energy splitting must be lower than the superconducting energy gap[1]. This is not necessarily the case for superconductors that lack a center of symmetry (NCS), however. The absence of inversion symmetry, when present in systems together with spin orbit coupling (SOC), introduces an antisymmetric spin-orbit coupling (ASOC)[2,3] term into the description of the electronic system that leads to a splitting of electronic bands. As a result, a mixture of singlet and triplet pairing can be observed[2,4–6] and the upper critical field can potentially be larger than predicted by the Pauli relation. For this reason, NCS superconductors are of significant interest[3,7–15].

Here we describe two previously unreported compounds, NbIr$_2$B$_2$ and TaIr$_2$B$_2$, the first known ternary compounds in the Nb-Ir-B and Ta-Ir-B chemical systems. They have a previously unreported, low symmetry *Cc* noncentrosymmetric crystal structure. Magnetization, heat capacity and resistivity measurements confirm presence of superconductivity with $T_c$'s = 7.2 K and 5.1 K. The estimated upper critical fields $\mu_0H_{c2}(0)$ = 16.3 T and 14.7 T respectively, significantly exceed the Pauli limit (which for NbIr$_2$B$_2$ is 13.3 T and for TaIr$_2$B$_2$ is 9.5 T). Our electronic band structure calculations show that the Fermi surface is mostly formed by Ir-5$d$ orbitals and is split by strong SOC. The theoretical results support a multigap scenario for NbIr$_2$B$_2$ – which we speculate to be present based on the analysis of the heat capacity data in the superconducting state of that material.

## 2. Results and discussion

### 2.1. Crystal structure

NbIr$_2$B$_2$ adopts a previously unreported structure type in the space group *Cc* (no. 9), determined by single crystal X-ray diffraction (see Table S1 in the Supplementary Material (SM)), shown in Figure 1(a). Details of the crystal structure, i.e. atomic coordinates and anisotropic thermal displacements are provided in Table S2 and Table S3, respectively. The single crystal structure determination shows that Boron dimers occupy the voids in the five edge-sharing Nb@Ir$_9$ polyhedra (see Figure 1(a)). NbIr$_2$B$_2$ and TaIr$_2$B$_2$ are isoelectronic to noncentrosymmetric superconductors NbRh$_2$B$_2$ and TaRh$_2$B$_2$, which are found in the chiral space group *P*3$_1$, instead of the current monoclinic space group[9]. TT'$_2$B$_2$ (T= Nb and Ta; T'= Rh and Ir) share common structural features, as shown in Figure 1(a). Two repeating units are present in TIr$_2$B$_2$, labelled as X and Y. There is a third type of repeating unit, marked as Z, found in TRh$_2$B$_2$. Note that the difference between Y and Z is that the B dimers are not aligned in parallel. Therefore, TIr$_2$B$_2$ can be interpreted as a stacking system with a pattern of XYXYXY while TRh$_2$B$_2$ stacks as XYZXYZXYZ. The difference between Ir and Rh atoms plays an important role in determining that the stability of the repeating unit Z. Figure 1(b) shows coordination of two distinct boron sites (B4 and B5) in NbIr$_2$B$_2$ to different atoms (Nb3, Ir1 and Ir2). One can find that Nb3-B4 and Ir1-B5 construct edge-shared distorted six-member ring layers while Nb3-B5 and Ir1-B4 build up six-member helical rings. In these four frameworks, boron atoms are three coordinated to Nb3/Ir1 atoms. While turning to the other Ir site, marked as Ir2, boron atoms become two-coordinated with Ir atoms and construct quasi-one-dimensional distorted Ir-B zig-zag chains.

We also performed powder X-ray diffraction (pXRD) on powderized NbIr$_2$B$_2$ and TaIr$_2$B$_2$. Rietveld refinements of the pXRD diffraction patterns (Figure S1) confirm that both compounds crystallize in the same monoclinic, noncentrosymmetric structure and show that replacement of the 4$d$ element Nb by the 5$d$ element Ta causes a small decrease (approximately 0.5 %) of the unit cell volume.

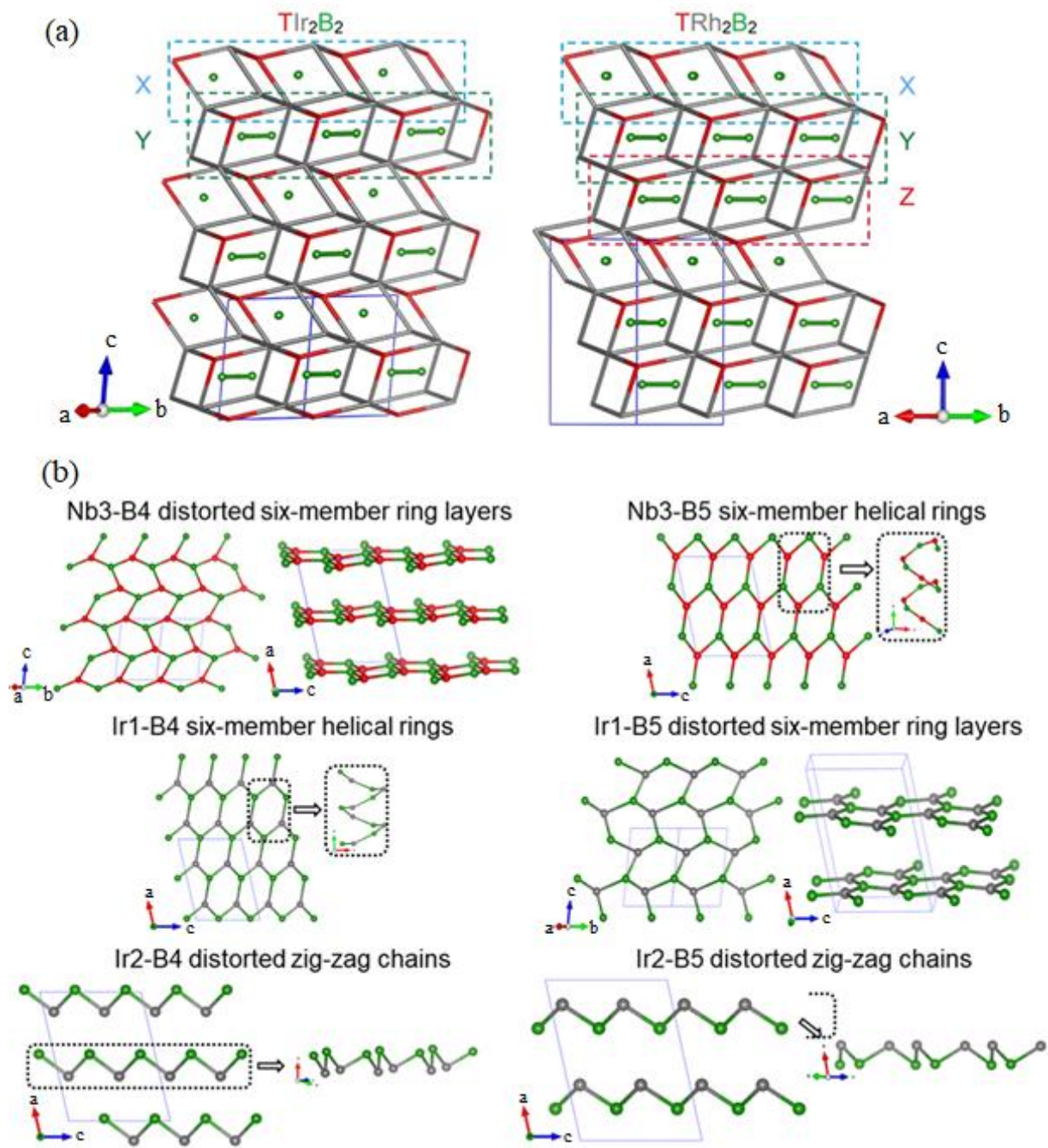

**Figure 1**: (a) The structural comparison between TIr$_2$B$_2$ and TRh$_2$B$_2$ with emphasis on the stacking pattern difference. (b) The coordination environments of the Boron atoms in TIr$_2$B$_2$.

The EDS (Energy Dispersive X-ray Spectroscopy) analysis of these materials reveals that the Nb(Ta):Ir ratio, 1:2, is consistent with the nominal composition (Nb(Ta)Ir$_2$B$_2$), confirming the refined structural model (for the Nb(Ta) and Ir, the B content is not quantitatively determined by this method).

**2.2. Superconducting properties**

The superconductivity in NbIr$_2$B$_2$ and TaIr$_2$B$_2$ is revealed through the magnetic measurements shown in Figure 2(a)-(d). Panels (a) and (b) present the temperature dependence of the volume dc magnetic susceptibility $\chi_V(T)$ with a clear transition to the superconducting state. Superconducting critical temperature determined from the magnetic susceptibility is estimated as the point at which the line set by the steepest slope of the superconducting signal in the zero-field cooled data set intersects with the extrapolation of the normal-state magnetic susceptibility to lower temperatures[16]. The critical temperature is estimated to be $T_c$ = 7.2 K and $T_c$ = 5.1 K for NbIr$_2$B$_2$ and TaIr$_2$B$_2$, respectively. Correcting the dc susceptibility data for the demagnetization factor (derived from M($\mu_0$H) studies as it is described in the SM), N = 0.49 for the Nb variant and N = 0.55 for the Ta variant, the ZFC measurements are consistent with 100% Meissner volume fraction. The N values are fairly consistent with the expected (theoretical) $N_z$ value derived for a circular cylinder sample with the height to radius ratio of approx. 0.5 (see ref.[17]). Compared with the ZFC data, the observed FC signal is much weaker, which is typical for polycrystalline samples.

Discussing the characterization of the superconductors in more detail, the magnetization versus applied magnetic field M($\mu_0$H) curves over a range of temperatures below the superconducting critical temperature are shown in the inset of Figure 2(c) and Figure 2(d). The first deviation from linearity from the initial slope is taken as the basis to determine the value of the lower critical field ($\mu_0 H_{c1}^*$) in these type-II superconductors. In order to precisely calculate this point, and also obtain a demagnetization factor N, we follow the methodology described in the SM and in the literature[12,18]. The resulting estimated values of $\mu_0 H_{c1}^*$ are depicted in the main panel of Figure 2(c) and Figure 2(d). An additional point for H = 0 is the zero field transition temperature taken from the resistivity measurement. The data points are modeled using the expression:

$$\mu_0 H_{c1}^*(T) = \mu_0 H_{c1}^*(0)\left[1 - \left(\frac{T}{T_c}\right)^n\right] \tag{1}$$

where $\mu_0 H_{c1}^*(0)$ is the lower critical field at 0 K and $T_c$ is the superconducting critical temperature. A typical $\mu_0 H_{c1}(T)$ relation has parabolic character (n=2) although there is no fundamental significance of the parabolic shape[19]. Our experimental data are well described with the above formula and the fit (red solid line) yields n = 3.8(3) and $\mu_0 H_{c1}^*(0)$ = 6.5(1) mT for the Nb variant, and n = 2.5(3) and $\mu_0 H_{c1}^*(0)$ = 2.71(5) mT for the Ta variant. The values for other two refined parameters are: $\mu_0 H_{c1}^*(0)$ = 6.5(1) mT and $T_c$ = 7.23(5) K for Nb variant, $\mu_0 H_{c1}^*(0)$ = 2.71(5) mT and $T_c$ = 5.41(6) K for Ta variant. Taking into account the demagnetization factor N for each sample, the lower critical field ($\mu_0 H_{c1}$) at 0 K was calculated from the formula:

$$\mu_0 H_{c1}(0) = \mu_0 H_{c1}^*(0)/(1-N)$$

The obtained values are $\mu_0 H_{c1}^*(0)$ = 13.0 mT for NbIr$_2$B$_2$ and $\mu_0 H_{c1}^*(0)$ = 6.0 mT for TaIr$_2$B$_2$.

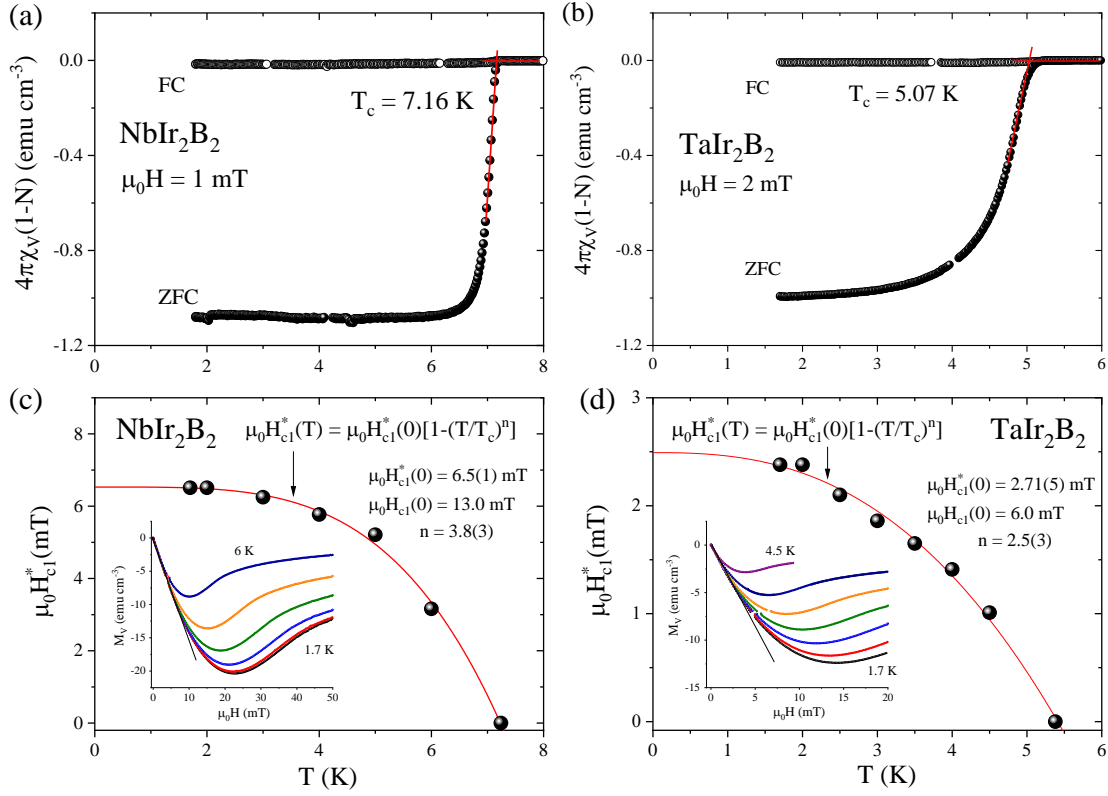

**Figure 2.** Temperature dependences of the zero-field-cooled (ZFC) and field-cooled (FC) volume magnetic susceptibility measured in a magnetic field of 1mT for NbIr$_2$B$_2$ (a) and 2mT for TaIr$_2$B$_2$ (b). The temperature dependence of the lower critical fields for NbIr$_2$B$_2$ (c) and TaIr$_2$B$_2$ (d). The inset shows the field-dependent magnetization curves $M_V(H)$ taken at different temperatures.

The low-temperature heat capacity measurements were performed to confirm the bulk nature of the superconductivity and in order to obtain important superconducting parameters, i.e. the normalized specific heat jump ($\Delta C/\gamma T_c$) and electron-phonon coupling constant ($\lambda_{e-p}$). The results are shown in Figure 3(a)-(c). A pronounced, large anomaly in the zero-field $C_p/T$ data confirms the bulk nature of the superconductivity for both compounds. From the equal entropy construction (blue solid lines) one finds the critical temperature $T_c = 7.18$ K for the Nb-based compound (Figure 3(a)) and $T_c = 5.11$ K for the Ta-based compound (Figure 3(b)). The variations of $C_p/T$ with $T^2$ at lower temperature and under 8 T magnetic field are presented in the insets of Figure 3(a) and Figure 3(b). The normal state specific-heat data can be fitted using the equation $C_p/T = \gamma + \beta T^2$, where the first and second terms are attributed to the electronic and lattice contributions to $C_p$, respectively. The extrapolation gives $\gamma$ = 4.9(3) mJ mol$^{-1}$K$^{-2}$ and $\beta$ = 0.470(7) mJ mol$^{-1}$K$^{-4}$ for NbIr$_2$B$_2$ and $\gamma$ = 5.2(3) mJ mol$^{-1}$K$^{-2}$ and $\beta$ = 0.80(1) mJ mol$^{-1}$K$^{-4}$ for TaIr$_2$B$_2$. Having calculated the Sommerfeld coefficient ($\gamma$) and the specific heat jump ($\Delta C/T_c$) at $T_c$, another important superconducting parameter can be obtained. The normalized specific heat jump ($\Delta C/\gamma T_c$) is equal 2.94 and 1.44 for NbIr$_2$B$_2$ and TaIr$_2$B$_2$, respectively. In the case of Nb variant compound, the calculated value (2.94) is much larger than the expected value of 1.43 for the BCS weak coupling limit and suggests that strongly coupled electrons are involved in the superconductivity in this compound. Such a large value of $\Delta C/\gamma T_c$ was reported for Mo$_3$Al$_2$C (2.14)[20], W$_3$Al$_2$C (2.7)[7], KOs$_2$O$_6$ (2.87)[21], Rh$_{17}$S$_{15}$ (2.0)[22] or IrGe (3.04)[23].

In the next step, the Debye temperature $\Theta_D$ was calculated using the relation:

$$\Theta_D = \left(\frac{12\pi^4}{5\beta}nR\right)^{1/3} \quad (2)$$

where R = 8.31 J mol$^{-1}$K$^{-1}$ is a gas constant, n = 5 is the number of atoms per formula unit. The values of $\Theta_D$ were estimated to be 274(1) K for NbIr$_2$B$_2$ and 230(1) K for TaIr$_2$B$_2$. Having the calculated Debye temperature $\Theta_D$, the electron-phonon constant $\lambda_{e-p}$, a dimensionless number that describes the coupling between the electron and the phonon, can be calculated from the inverted McMillan formula[24]:

$$\lambda_{e-p} = \frac{1.04 + \mu^* ln(\Theta_D/1.45\,T_c)}{(1 - 0.62\mu^*)ln(\Theta_D/1.45\,T_c) - 1.04} \quad (3)$$

where $\mu^*$ is the Coulomb pseudopotential parameter having typical material specific values in the range 0.1 $\leq \mu^* \leq$ 0.15, where 0.13 is typically used for intermetallic superconductors[9,25–27]. The constant $\lambda_{e-p}$ = 0.74 for NbIr$_2$B$_2$ and $\lambda_{e-p}$ = 0.70 for TaIr$_2$B$_2$, suggesting that both compounds are moderately or strongly coupled superconductors.

The temperature dependence of the electronic specific heat ($C_{el}$) below $T_c$ for NbIr$_2$B$_2$ is shown in Figure 3(c). The $C_{el}$ was then analyzed by fitting the data with a single gap isotropic *s*-wave model and an isotropic two-gap (*s+s*)-wave model. Both fits were done below 5K, which is about 0.7$T_c$ and the expected by BCS theory value energy gap is $2\Delta_0 = 3.52k_BT_c = 2.17$ meV. An *s*-wave single gap BCS model (blue dashed line) gives $2\Delta_0 = 2.70(6)$ meV. A better fit was obtained assuming a multigap (*s+s*) scenario with $2\Delta_{01} = 2.32(5)$ meV and $2\Delta_{02} = 9.1(12)$ meV, represented by a red line. The dashed and solid lines in the inset represent the difference between the experiment and a single and double *s*-wave gap model, respectively. For a gap with nodes, theory predicts power-law dependence, which does not work here; the fits are shown in Figure S3 of the SM. More experiments that shed light on the gap symmetry are required. For example, multigap superconductivity, probed by the μSR technique, has been reported for isoelectronic but not isostructural TaRh$_2$B$_2$[28].

The last experimental technique used for characterization of the new superconductors was temperature dependent resistivity with the results shown in the main panel of Figure 3(d) for NbIr$_2$B$_2$ and 3(e) for TaIr$_2$B$_2$. NbIr$_2$B$_2$ behaves like a poor metal, with a shallow negative gradient for the resistivity upon cooling from room temperature. The residual resistivity ratio (RRR), $\rho_{300}/\rho_{10} = 1.3$, is small, which can be attributed to the polycrystalline nature of the sample contained grain boundaries and macroscopic defects. In the case of Ta variant, one observes an increase in $\rho(T)$ as the temperature was decreased. Comparing $\rho(300$ K$)$ and $\rho(10$ K$)$, resistivity increases about 50%. This behavior could be due to a weak localization (WL) of charge carriers due to disorder. The experimental data were fitted with the function[29,30] $\rho(T) = [1/(\sigma_0 + \alpha T^{\frac{p}{2}})] + \beta T$, where $\sigma_0 = 1/\rho_0$ is the residual conductivity, p is related to the temperature dependence of the inelastic scattering time and the second term describes the high temperature part. The experimental data are very well described with this model ($R^2=0.9996$), yielding the fit parameters $\sigma_0 = 9.02(1)\times10^{-4}$ μΩ$^{-1}$cm$^{-1}$, $\alpha = 6.5(2)\times10^{-4}$ μΩ$^{-1}$cm$^{-1}$K$^{-p/2}$, p = 2.92(2) and $\beta = 1.71(2)$ μΩcmK$^{-1}$.

At low temperatures the electrical resistivity drops sharply to zero at $T_c = 7.24$ K for NbIr$_2$B$_2$ and at $T_c = 5.38$ K for TaIr$_2$B$_2$, where $T_c$ is defined as the midpoint of the transition. The slightly higher superconducting temperature value obtained in the resistivity measurement is likely due to the influence of surface superconductivity emerging in each cross-sectional area of the sample. The effect of applying a magnetic field on $T_c$ is shown in the insets of Figure 3(d) and Figure 3(e). As expected, the transition becomes broader, and $T_c$ shifts to lower temperature, as the applied field is increased. It should be noted that a transition to a zero-resistance state was obtained even at 9 T and above 3 K for NbIr$_2$B$_2$ or 2 K for TaIr$_2$B$_2$, indicating a large upper critical field.

Using the midpoint resistivity, the upper critical field ($\mu_0H_{c2}$) for both compounds, plotted as a function of temperature, is illustrated in Figure 3(f). For TaIr$_2$B$_2$ one observes a small concave-upwards curvature curve of $\mu_0H_{c2}$ versus T. Such behavior is a typical feature observed for conventional superconductors with an anisotropic Fermi surface, and has been observed in multigap superconductors, as well as in unconventional superconductors[22]. The solid line, presented in Figure 3(f), is a fit to the Ginzburg-Landau expression:

$$\mu_0H_{c2}(T) = \mu_0H_{c2}(0)\frac{(1-t^2)}{(1+t^2)} \qquad (4)$$

where $t = T/T_c$ and $T_c$ is the transition temperature at zero magnetic field. Our experimental data fit Equation (4) fairly well. The obtained values of $\mu_0H_{c2}(0)$ are: 16.3(2) T and 14.7(1) T for NbIr$_2$B$_2$ and TaIr$_2$B$_2$, respectively. According to the BCS theory, the Pauli-limiting field can be obtained from $\mu_0H_{c2}^p(0) = 1.85T_c$, which for NbIr$_2$B$_2$ gives $\mu_0H_{c2}^p(0) = 13.3(1)$ T and $\mu_0H_{c2}^p(0) = 9.5(1)$ T for TaIr$_2$B$_2$. The experimentally estimated $\mu_0H_{c2}(0)$ values obtained for the current materials are roughly 20% and 50% larger than the $\mu_0H_{c2}^p$, and hence suggest that the materials may exhibit non-BCS superconductivity. The critical temperatures extracted from the anomaly in $C_p(T)$ at the superconducting transition are also added to Figure 3(f) (filled circles and squares). The thermodynamic data were fitted with Equation (4) (dashed line), yielding $\mu_0H_{c2}(0) = 15.8(1)$ T for Nb variant and $\mu_0H_{c2}(0) = 16.5(2)$ T for Ta variant. Table S7 (SM) gathers $\mu_0H_{c2}(0)$ values obtained from GL and WHH models. In all cases the $\mu_0H_{c2}(0)$ exceeds the Pauli-limiting field.

Consequently, the characteristic Ginzburg-Landau coherence length, $\xi_{GL}$, can be obtained using the relation

$$\mu_0 H_{c2}(0) = \frac{\Phi_0}{2\pi \xi_{GL}^2}, \tag{5}$$

where $\Phi_0 = hc/2e$ is the quantum flux, where $\mu_0 H_{c2}$ were taken from the GL fit to the resistivity data. This way, the value of $\xi_{GL}$ was estimated to be 45 Å for $NbIr_2B_2$ and 47 Å for $TaIr_2B_2$. In the next step, the Ginzburg-Landau penetration depth $\lambda_{GL}(0)$ can be calculated from the relation

$$\mu_0 H_{c1}(0) = \frac{\Phi_0}{4\pi \lambda_{GL}^2} \ln\frac{\lambda_{GL}}{\xi_{GL}}. \tag{6}$$

The value is found to be $\lambda_{GL}(0) = 2230$ Å for Nb variant and $\lambda_{GL}(0) = 3420$ Å for Ta variant. From the equation $\kappa_{GL} = \lambda_{GL}/\xi_{GL}$, the Ginzburg-Landau parameter $\kappa_{GL}$ is about 50 for $NbIr_2B_2$ and 72 for $TaIr_2B_2$ and therefore, it is clear that each superconducting material is a type-II superconductor ($\kappa_{GL} > 1/\sqrt{2}$). Finally, the thermodynamic critical field can be obtained from $\kappa_{GL}$, $H_{c1}$ and $H_{c2}$ using the formula

$$H_{c1} H_{c2} = H_c^2 \ln\kappa_{GL} \tag{7}$$

yielding $\mu_0 H_c = 232$ mT for $NbIr_2B_2$ and $\mu_0 H_c = 144$ mT for $TaIr_2B_2$. All the superconducting and normal state parameters of $NbIr_2B_2$ and $TaIr_2B_2$ are gathered in Table I.

**Table 1.** Superconducting parameters of TIr$_2$B$_2$ where T = Nb and Ta.

| Parameter | Unit | NbIr$_2$B$_2$ | TaIr$_2$B$_2$ |
|---|---|---|---|
| $T_c$ | K | 7.18 | 5.11 |
| $\mu_0 H_{c1}(0)$ | mT | 13.0 | 6.0 |
| $\mu_0 H_{c2}(0)$ | T | 16.3 | 14.7 |
| $\mu_0 H^{Pauli}$ | T | 13.3 | 9.5 |
| $\xi_{GL}$ | Å | 45 | 47 |
| $\lambda_{GL}$ | Å | 2230 | 3420 |
| $\kappa_{GL}$ | --- | 50 | 72 |
| $\gamma$ | mJ mol$^{-1}$ K$^{-2}$ | 4.9 | 5.2 |
| $\Delta C_p/\gamma T_c$ | --- | 2.94 | 1.44 |
| $\lambda_{e-p}$ | --- | 0.74 | 0.70 |
| $\Theta_D$ | K | 274 | 230 |

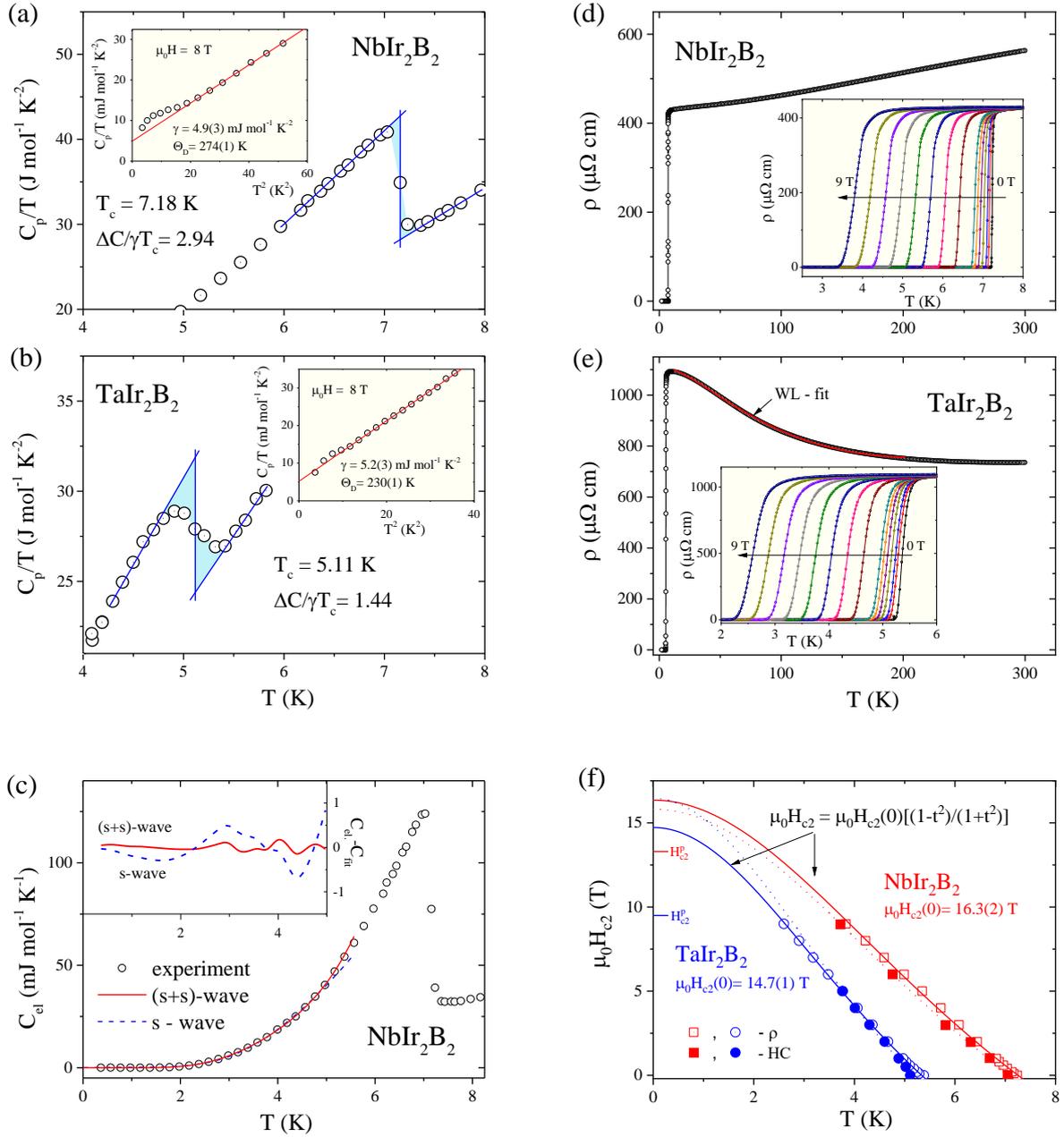

**Figure 3.** Further characterization of the superconductors. The specific heat jump in zero magnetic field at low temperatures with $T_c = 7.18$ K for NbIr$_2$B$_2$ (a) and $T_c = 5.11$ K for TaIr$_2$B$_2$ (b). Inset: $C_p/T$ versus $T^2$ measured in 8-T field (in the normal state) fitted to $C_p/T = \gamma + \beta T^2$. (c) Temperature-dependent electronic specific heat $C_{el.}$ for NbIr$_2$B$_2$ with a fit of a single gap isotropic *s*-wave model (blue dashed line) and an isotropic two-gap (*s+s*)-wave model (red solid line) to the data. The temperature dependence of the electrical resistivity of NbIr$_2$B$_2$ (d) and TaIr$_2$B$_2$ (e) measured in zero magnetic field. The inset shows the low temperature resistivity data taken in several different magnetic fields. (f) Temperature dependence of the upper critical field, determined from the electrical resistivity (open points) and heat capacity (full points) data.

## 2.3. Electronic band structure

To get an insight into the electronic structure of our compounds, density functional theory (DFT) calculations were performed. Figure 4 shows the computed electronic bands E(**k**) and densities of states (DOS), whereas Fermi surface is displayed in Figure 5. As seen from the atomic character of DOS near $E_F$, the Fermi surface will be mainly formed from Nb-4$d$ (Ta-5$d$) and Ir-5$d$ orbitals, and due to larger atomic population, contribution from Ir is larger. Figure 4(a,c) shows that two bands are crossing $E_F$ in a scalar-relativistic case, and due to combination of spin-orbit coupling and lack of inversion symmetry, bands are split. The value of energy band splitting ($E_{ASOC}$) is strongly **k**-dependent and ranges from 25 meV to 250 meV, as large as is seen in CePt$_3$Si or Li$_2$Pt$_3$B[15] (~200 meV) and larger than in LaNiC$_2$[31,32], (~40 meV), where nonunitary triplet pairing has been proposed[31]. Generally, systems with large $E_{ASOC}$ are promising to look at for singlet-triplet mixing[15]. SOC has a negligible effect on the DOS($E_F$) in NbIr$_2$B$_2$, but, for TaIr$_2$B$_2$, the relativistic value is reduced by about 20% due to a shift in the DOS peak position. This is correlated with the smaller $T_c$ in this compound, which additionally enhances the ratio of $E_{ASOC}$ to $k_B T_c$, the most fundamental superconducting parameter correlated with the presence of an unconventional pairing symmetry[33].

In Figure 4(e,g) the band structure is projected on the spin directions, showing the mixed spin character.

Figure 5 shows the calculated Fermi surface (FS) and FS cross-sections for NbIr$_2$B$_2$ (panels a-j) and TaIr$_2$B$_2$ (panels k-u). Spin-orbit coupling not only splits each FS sheet into two pieces, but also the topology of the Fermi surface is strongly affected. Especially the second FS sheet, shown in panels (b) and (l), is visibly modified after introducing SOC, see panels (d,f) and (n,p) for Nb analog and Ta analog, respectively. The reason for such a strong modification of the FS is seen in the bandstructure plots in Figure 4 (a,c). Due to SOC the highest band (among those which cross $E_F$) is shifted towards higher energy and the number of points where this band crosses $E_F$ is reduced, leading to a smaller area of this FS sheet. FS cross-sections, shown in Figure 5 (g-j) and (r-u), additionally visualize the Fermi surface mismatch and observation, that SOC effect on FS in the studied materials is more than splitting of the Fermi surface into a set of similar parallel sheets.

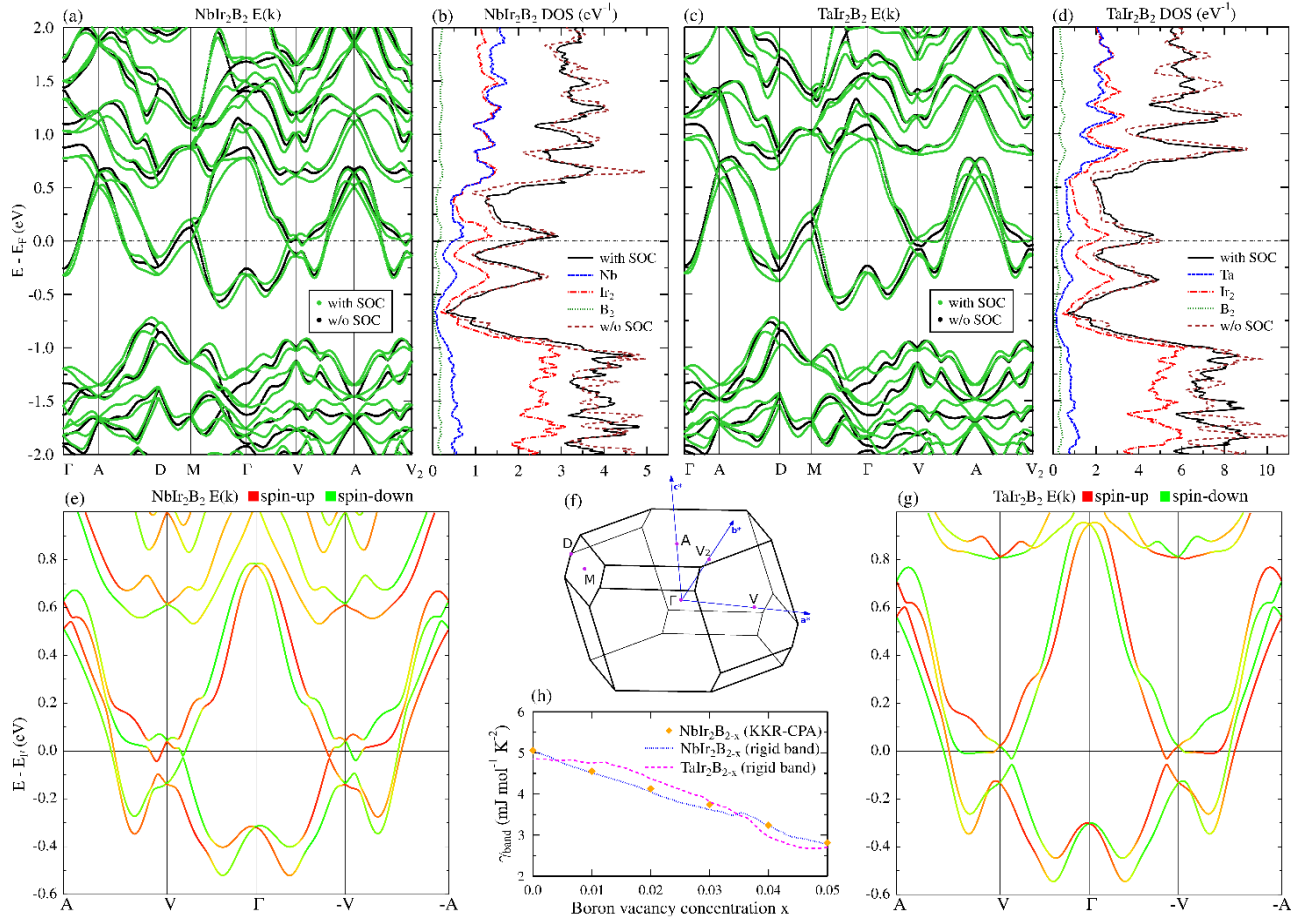

**Figure 4.** Calculated electronic structures of the superconductors. (a,c) electronic dispersion relations E(**k**), computed with and without spin-orbit coupling, band splitting due to SOC is clearly visible; (b,d) densities of states (DOS); (e,g) zoom of E(**k**) where spin character of each band is marked with color. Bands have a mixed, spin "up" - "down" character. In calculations monoclinic b-axis was chosen as a magnetization direction. Our compounds are non-magnetic materials, the time reversal symmetry is preserved thus in spin-split bands E(**k**) = E(-**k**) degeneracy is kept, however spin direction is flipped, when **k** is changed to -**k**; (f) the Brillouin zone; (h) evolution of the computed value of the Sommerfeld parameter $\gamma_{band}$ with the boron vacancy concentration x in {Nb,Ta}Ir$_2$B$_{2-x}$. KKR-CPA calculations were performed for NbIr$_2$B$_{2-x}$ for x = 0, 0.01, 0.02, 0.03, 0.04 and 0.05, and perfectly follow the rigid band model prediction, where $E_F$ is shifted in the stoichiometric x = 0 DOS according to the reduced number of electrons in the system.

Returning to the densities of states, the computed band structure DOS($E_F$) values (with SOC included) are equal to 2.14 eV$^{-1}$ (Nb analog) and 2.06 eV$^{-1}$ (Ta analog), which result in the band values of the Sommerfeld electronic specific heat parameter $\gamma_{band}$ = 5.05 mJ mol$^{-1}$ K$^{-2}$ for NbIr$_2$B$_2$ and $\gamma_{band}$ = 4.85 mJ mol$^{-1}$ K$^{-2}$ for TaIr$_2$B$_2$. The experimental values are equal to 4.9 mJ mol$^{-1}$ K$^{-2}$ and 5.2 mJ mol$^{-1}$ K$^{-2}$, respectively, which leads to a puzzling situation, since almost equal "bare" bandstructure and experimental $\gamma$ values leave no room for the electron-phonon renormalization factor, where $\gamma$ = $\gamma_{band}(1+\lambda_{e-p})$. The $\lambda_{e-p}$, estimated from the critical temperature via the McMillan formula is about 0.7 in both materials, thus we expect either a smaller $\gamma_{band}$ values, of the order of 3 mJ mol$^{-1}$ K$^{-2}$ for both compounds, or about 70% larger than the measured $\gamma$ values. As the accuracy of the measured $\gamma$ is certainly much better than 10%, other explanations must be considered. As we have shown in Figure 4(b,d) in both compounds E$_F$ is located at the steep DOS slope. If we assume that the studied samples are slightly electron-deficient, e.g. due to the formation of boron vacancies, we may explain the discrepancy in the $\gamma$ values, as in such situation E$_F$ will move to the lower energies, considerably decreasing DOS(E$_F$) values. Quantitatively this analysis is shown in Figure 4(h). Boron vacancies are expected to rigidly shift E$_F$ downwards, with each vacancy delivering 3 holes to the system, as boron is a trivalent element. To reach $\gamma_{band}$ ~3 mJ mol$^{-1}$ K$^{-2}$ it is sufficient to assume having at most 2.5 % of boron deficiency (i.e. {Nb,Ta}Ir$_2$B$_{1.95}$). To cross-check the computed $\gamma_{band}$ values and verify the assumption of rigid-band-like behavior in the boron-deficient system, additional calculations were done for NbIr$_2$B$_{2-x}$. We used the Korringa-Kohn-Rostoker method, and the presence of boron vacancies was explicitly taken into account using the coherent potential approximation[34,35]. KKR-CPA calculations confirmed that boron vacancy rigidly shifts Fermi level position leading to the decrease in DOS(E$_F$) value, see Figure S4. As shown in Figure 4(h) the x = 0 value of $\gamma_{band}$ obtained from KKR-CPA perfectly agrees with the one obtained using FP-LAPW. Thus, the observed discrepancy in the Sommerfeld parameter values suggests that a small amount of B vacancies are present in the system.

Such a small boron deficiency (2.5%) is certainly not possible to detect by EDS or pXRD technique. It is worth noting that for the MgCNi$_3$ superconductor, a powder neutron diffraction analysis reveals that the carbon occupancy is 0.978(8), though 25% excess of carbon has been used in the synthesis[36]. Hence, similar situation might occur in preparation of {Nb,Ta}Ir$_2$B$_2$.

If we accept the hypothesis, that $E_F$ in the studied materials is rigidly shifted to lower energies, the required shift to match the experimental and renormalized calculated Sommerfeld parameter is equal to 85 meV (NbIr$_2$B$_2$) and 73 meV (TaIr$_2$B$_2$). In such a case the contribution of the first two FS sheets (Figure 5(c,e,m,o)) to the total Fermi surface will increase, limiting the role of the third (Figure 5(d,n)) and especially the fourth (Figure 5(f,p)) one in electronic structure of the materials. The asymmetry between the last two FS sheets will also be larger. The presence of two dominating FS sheets fits in with the hypothesis of two superconducting gaps. Additional Fermi surface plots for the shifted $E_F$ are presented in Figures S5 and S6.

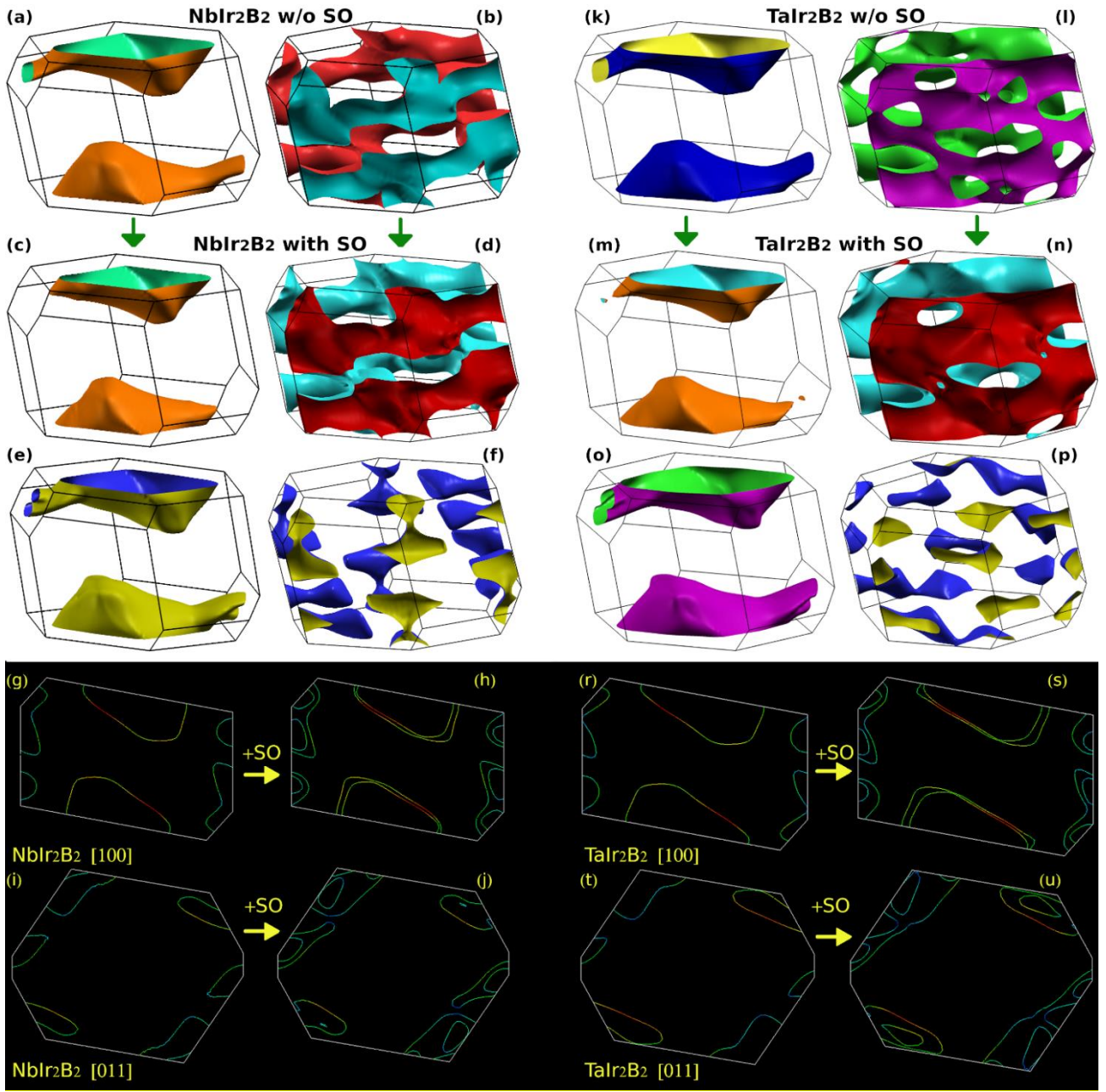

**Figure 5.** Impact of spin-orbit (SO) interaction on the Fermi surface (FS) of $NbIr_2B_2$ and $TaIr_2B_2$. In both compounds, two FS sheets [panels (a,b) and (k,l)] are split into four sheets [panels (c-f) and (m-p)]. FS splitting is well visible on FS cross sections, shown in panels (g-j) and (r-u) for $NbIr_2B_2$ and $TaIr_2B_2$, respectively.

## 3. Conclusions

The first rule proposed by Matthias and Hulm for superconductors was: "high symmetry is good, cubic symmetry is best"[37]. History, through the copper oxide and iron pnictide superconductors, has taught us that this is not generally the case, but this rule can still be imagined to hold for conventional intermetallic superconductors. The two intermetallic compounds reported here, which display a previously unreported noncentrosymmetric low symmetry crystal structures appear to violate that rule. The new compounds are type-II superconductors with $T_c$ = 7.2 K and 5.1 K for $NbIr_2B_2$ and $TaIr_2B_2$. Unlike the well-known boron-based superconductor $MgB_2$[38] and heavy-fermion superconductor $YbAlB_4$[39] where boron atoms form isolated layers, i.e., honeycomb layer for $MgB_2$ and edge-shared five-member ring with seven-member ring for $YbAlB_4$, the boron dimers present in noncentrosymmetric $TT'_2B_2$ superconductors may lead to a novel design rule for boron-based superconductors.

The most important common characteristic of both families of noncentrosymmetric $TT'_2B_2$ superconductors is their large value of the upper critical field. For the current materials, the estimated upper critical fields are $\mu_0H_{c2}$ = 16.3 T and 14.7 T, both of which exceed their so-called Pauli limits. Analysis of heat capacity data in the superconducting state for $NbIr_2B_2$ suggests a possible 2 gap (*s+s*) pairing symmetry function. In noncentrosymmetric compounds, the degree of admixture of spin-singlet and spin-triplet states in the superconductor depends on the strength of the spin orbit coupling. $NbIr_2B_2$ $TaIr_2B_2$ therefore appear to form a good family for investing the impact of atomic make up on the degree of spin orbit coupling at the Fermi surface and its effect on superconductivity.

4. **Experimental**

The starting materials for the synthesis of $NbIr_2B_2$ and $TaIr_2B_2$ were elemental niobium (3N, 200 mesh, Sigma-Aldrich), tantalum (3N, 100 mesh, Alfa Aesar), iridium (4N, Mennice – Metale, Poland) and boron (submicron particles, Callery Chemical). Powders of Nb/Ta, Ir and B were weighed out in a 1:2:2.33 ratio, ground thoroughly using a mortar and pestle and pressed into a pellet using a hydraulic press. The samples (~200 mg) were then wrapped in tantalum foil, placed in an alumina crucible and heat treated at 1100°C for 13 hours under high vacuum ($10^{-6}$ torr). Mass loss during the synthesis was negligible.

Multiple crystals (of size ~30×30×30 μm$^3$) were measured at 300 K to get precise structural information. A Bruker Apex II diffractometer equipped with Mo radiation ($\lambda_{K\alpha}$= 0.71073 Å) was employed at room temperature. The small crystals were stuck to a Kapton loop with glycerol. Ten different detector positions were chosen to take the diffraction intensity measurements with an exposure time of 20 seconds per frame and a scanned 2θ width of 0.3°. Direct methods and full-matrix least-squares on F$^2$ within the SHELXTL package were employed to solve the structure[40]. Lorentz and polarization effects were modeled by the SAINT program, and numerical absorption corrections were accomplished with *XPREP*, which is based on face-index modeling[41]. Powder X-ray diffraction analysis on well-ground powder of a portion of samples was carried out on a Bruker D8 Advance Eco diffractometer with Cu Kα radiation and a LynxEye-XE detector. Having the crystallographic data of new compounds, Rietveld refinement of this data was performed by employing the software Topas. The Nb:Ir or Ta:Ir ratio was examined using a FEI Quanta 250 FEG scanning electron microscope (SEM) equipped with an Apollo-X SDD energy-dispersive spectrometer (EDS). The data were collected for 300 seconds and analyzed using the EDAX TEAM$^{TM}$ software.

Magnetization and magnetic susceptibility measurements were performed using a Physical Property Measurement System (Quantum Design PPMS) with a vibrating sample magnetometer (VSM) in the temperature range of 1.7 – 8.0 K under various applied magnetic fields. Both specific heat and electrical resistivity were measured in the temperature range between 300 and 1.85 K, in magnetic fields up to 9 T in the PPMS. The lower temperature heat capacity of NbIr$_2$B$_2$ was measured in a Dynacool Physical Property Measurement System equipped with a $^3$He attachment. The resistivity was determined using a standard four-probe technique, with four 37-$\mu$m-diameter platinum wire leads spark-welded to the flat polished sample surface. Specific-heat measurements were performed using the two-τ time-relaxation method. The sample was attached to the measuring platform by a small amount of Apiezon N. The addendum heat capacity was measured in a separate run without a sample and was subtracted from the data.

The electronic structure was calculated using the full-potential linearized augmented plane wave method (FP-LAPW) implemented in the WIEN2k package[42], using the Perdew-Burke-Ernzerhof generalized gradient approximation[43] (PBE-GGA) for the exchange-correlation potential. Calculations were done using the experimental lattice parameters, and for both the experimental and computed (relaxed) atomic positions, however the relaxation process does not lead to any visible changes in the calculated electronic band structure. Calculations were done in a scalar-relativistic (spin-orbit interaction is neglected) and relativistic (spin-orbit interaction included) way. Fermi surface plots and FS cross-sections were prepared using XCrysDen[44] and FermiSurfer[45] software. To simulate the effect of boron vacancies on the DOS($E_F$), the Korringa-Kohn-Rostoker method with the coherent potential approximation[34,35] was applied.

## Acknowledgments

The work at GUT was supported by the National Science Centre (Poland), grant number: UMO-2017/27/B/ST5/03044. Weiwei Xie was supported by Beckman Young Investigator Program and NSF-DMR-1944965. Work at AGH was supported by the National Science Centre (Poland), project No. 2017/26/E/ST3/00119. The synthetic work at Princeton was supported by the US Department of Energy, Basic Energy Sciences, grant DE-FG02-98ER45706.

[1]   M. Tinkham, *Introduction to Superconductivity*, Courier Corporation, **1996**.
[2]   L. P. Gor'kov, E. I. Rashba, *Phys. Rev. Lett.* **2001**, *87*, 037004.
[3]   D. Singh, S. K. P., J. A. T. Barker, D. M. Paul, A. D. Hillier, R. P. Singh, *Phys. Rev. B* **2018**, *97*, 100505.
[4]   E. Bauer, G. Hilscher, H. Michor, C. Paul, E. W. Scheidt, A. Gribanov, Y. Seropegin, H. Noël, M. Sigrist, P. Rogl, *Phys. Rev. Lett.* **2004**, *92*, 027003.
[5]   P. A. Frigeri, D. F. Agterberg, A. Koga, M. Sigrist, *Phys. Rev. Lett.* **2004**, *93*, 099903.
[6]   E. Bauer, M. Sigrist, *Non-Centrosymmetric Superconductor:Introduction and Overview*, Springer-Verlag, Heidelberg, **2012**.
[7]   T. P. Ying, Y. P. Qi, H. Hosono, *Phys. Rev. B* **2019**, *100*, 094522.
[8]   S. K. P., D. Singh, P. K. Biswas, G. B. G. Stenning, A. D. Hillier, R. P. Singh, *Phys. Rev. Mater.* **2019**, *3*, 104802.
[9]   E. M. Carnicom, W. Xie, T. Klimczuk, J. Lin, K. Górnicka, Z. Sobczak, N. P. Ong, R. J. Cava, *Sci. Adv.* **2018**, *4*, eaar7969.
[10]  A. Amon, E. Svanidze, R. Cardoso-Gil, M. N. Wilson, H. Rosner, M. Bobnar, W. Schnelle, J. W. Lynn, R. Gumeniuk, C. Hennig, G. M. Luke, H. Borrmann, A. Leithe-Jasper, Y. Grin, *Phys. Rev. B* **2018**, *97*, 014501.
[11]  Y. Qi, Z. Xiao, J. Guo, H. Lei, T. Kamiya, H. Hosono, *EPL Europhys. Lett.* **2018**, *121*, 57001.
[12]  J. A. T. Barker, B. D. Breen, R. Hanson, A. D. Hillier, M. R. Lees, G. Balakrishnan, D. M. Paul, R. P. Singh, *Phys. Rev. B* **2018**, *98*, 104506.
[13]  A. Ptok, K. Domieracki, K. J. Kapcia, J. Łażewski, P. T. Jochym, M. Sternik, P. Piekarz, D. Kaczorowski, *Phys. Rev. B* **2019**, *100*, 165130.
[14]  I. Bonalde, H. Kim, R. Prozorov, C. Rojas, P. Rogl, E. Bauer, *Phys. Rev. B* **2011**, *84*, 134506.
[15]  M. Smidman, M. B. Salamon, H. Q. Yuan, D. F. Agterberg, *Rep. Prog. Phys.* **2017**, *80*, 036501.
[16]  T. Klimczuk, R. J. Cava, *Phys. Rev. B* **2004**, *70*, 212514.
[17]  M. Sato, Y. Ishii, *J. Appl. Phys.* **1989**, *66*, 983.
[18]  A. Umezawa, G. W. Crabtree, J. Z. Liu, T. J. Moran, S. K. Malik, L. H. Nunez, W. L. Kwok, C. H. Sowers, *Phys. Rev. B* **1988**, *38*, 2843.
[19]  A. C. Rose-Innes, E. H. Rhoderick, *Introduction to Superconductivity*, Pergamon Place Plc, **1978**.
[20]  E. Bauer, G. Rogl, X.-Q. Chen, R. T. Khan, H. Michor, G. Hilscher, E. Royanian, K. Kumagai, D. Z. Li, Y. Y. Li, R. Podloucky, P. Rogl, *Phys. Rev. B* **2010**, *82*, 064511.
[21]  Z. Hiroi, S. Yonezawa, Y. Nagao, J. Yamaura, *Phys. Rev. B* **2007**, *76*, 014523.
[22]  H. R. Naren, A. Thamizhavel, A. K. Nigam, S. Ramakrishnan, *Phys. Rev. Lett.* **2008**, *100*, 026404.
[23]  D. Hirai, M. N. Ali, R. J. Cava, *J. Phys. Soc. Jpn.* **2013**, *82*, 124701.
[24]  W. L. McMillan, *Phys. Rev.* **1968**, *167*, 331.
[25]  K. Górnicka, E. M. Carnicom, S. Gołąb, M. Łapiński, B. Wiendlocha, W. Xie, D. Kaczorowski, R. J. Cava, T. Klimczuk, *Supercond. Sci. Technol.* **2019**, *32*, 025008.
[26]  V. Y. Verchenko, A. A. Tsirlin, A. O. Zubtsovskiy, A. V. Shevelkov, *Phys. Rev. B* **2016**, *93*, 064501.
[27]  D. Singh, A. D. Hillier, A. Thamizhavel, R. P. Singh, *Phys. Rev. B* **2016**, *94*, 054515.
[28]  D. A. Mayoh, A. D. Hillier, K. Götze, D. M. Paul, G. Balakrishnan, M. R. Lees, *Phys. Rev. B* **2018**, *98*, 014502.
[29]  P. A. Lee, T. V. Ramakrishnan, *Rev. Mod. Phys.* **1985**, *57*, 287.
[30]  D. Vollhardt, P. Wolfle, *Electronic Phase Transitions*, Elsevier Science Publishers B.V, **1992**.
[31]  G. Csire, B. Újfalussy, J. F. Annett, *Eur. Phys. J. B* **2018**, *91*, e2018.
[32]  B. Wiendlocha, R. Szczęśniak, A. P. Durajski, M. Muras, *Phys. Rev. B* **2016**, *94*, 134517.
[33]  L. Jiao, J. L. Zhang, Y. Chen, Z. F. Weng, Y. M. Shao, J. Y. Feng, X. Lu, B. Joshi, A. Thamizhavel, S. Ramakrishnan, H. Q. Yuan, *Phys. Rev. B* **2014**, *89*, 060507.
[34]  H. Ebert, D. Ködderitzsch, J. Minár, *Rep. Prog. Phys.* **2011**, *74*, 096501.

[35] Ebert H., The Munich SPR-KKR package, version 7.7.3 http://ebert.cup.uni-muenchen.de/SPRKKR,**2018**.
[36] T. . Amos, Q. Huang, J. . Lynn, T. He, R. . Cava, *Solid State Commun.* **2002**, *121*, 73.
[37] K. Conder, *Supercond. Sci. Technol.* **2016**, *29*, 080502.
[38] J. Nagamatsu, N. Nakagawa, T. Muranaka, Y. Zenitani, J. Akimitsu, *Nature* **2001**, *410*, 63.
[39] S. Nakatsuji, K. Kuga, Y. Machida, T. Tayama, T. Sakakibara, Y. Karaki, H. Ishimoto, S. Yonezawa, Y. Maeno, E. Pearson, G. G. Lonzarich, L. Balicas, H. Lee, Z. Fisk, *Nat. Phys.* **2008**, *4*, 603.
[40] Sheldrick, G. M, *Acta Crystallogr. Sect. A* **2008**, *64*, 112.
[41] Sheldrick, G. M, *XPREP*, University of Gottingen, Gottingen, Germany, **2001**.
[42] P. Blaha, K. Schwarz, Madsen G. K. H., Kvasnicka D., Luitz J., *WIEN2K, An Augmented Plane Wave + Local Orbitals Program for Calculating Crystal Properties*, Technische Universitat Wien, **2001**.
[43] J. P. Perdew, K. Burke, M. Ernzerhof, *Phys. Rev. Lett.* **1996**, *77*, 3865.
[44] A. Kokalj, *Comput Mater Sci* **2003**, *28*, 155.
[45] M. Kawamura, *Comput. Phys. Commun.* **2019**, *239*, 197.

*Supplementary Material*

**SM-1. Single-crystal X-ray diffraction studies**

To determine the crystal structure of the new Nb-Ir-B phase, we carried out a single-crystal X-ray diffraction structure refinement at room temperature. A summary of the crystallographic data from the structure refinement, and the atomic coordinates, can be found in Tables S1 and S2 respectively. The anisotropic thermal displacements are gathered in Table S3. Since this method tests a micrometer size crystal, we also performed the powder X-ray (pXRD) diffraction for the samples used for physical properties characterization.

**Table S1**. Single crystal refinement for NbIr$_2$B$_2$ at 300 (2) K.

| **Refined Formula** | **NbIr$_2$B$_2$** |
|:---:|:---:|
| F.W. (g/mol) | 498.93 |
| Space group; Z | $Cc$; 4 |
| $a$ (Å) | 8.1586 (5) |
| $b$ (Å) | 4.7746 (3) |
| $c$ (Å) | 6.0067 (3) |
| $\beta$ (°) | 102.256 (3) |
| V (Å$^3$) | 228.65 (2) |
| Extinction Coefficient | 0.0029 (2) |
| θ range (deg) | 4.977-34.982 |
| R$_\sigma$ | 0.0563 |
| hkl range | $-13 \leq h \leq 12$<br>$-7 \leq k \leq 7$<br>$-9 \leq l \leq 9$ |
| No. reflections; R$_{int}$ | 3557; 0.0435 |
| No. independent reflections | 991 |
| No. parameters | 48 |
| $R_1$: $\omega R_2$ ($I>2\sigma(I)$) | 0.0254; 0.0433 |
| Goodness of fit | 0.885 |
| Diffraction peak and hole (e$^-$/ Å$^3$) | 2.359; -1.957 |
| Absolute structure parameter | 0.04 (2) |

**Table S2.** Atomic coordinates and equivalent isotropic displacement parameters of NbIr$_2$B$_2$ system. (U$_{eq}$ is defined as one-third of the trace of the orthogonalized U$_{ij}$ tensor (Å$^2$))

| Atom | Wyck. | Occ. | x | y | z | U$_{eq}$ |
|---|---|---|---|---|---|---|
| Ir1 | 4a | 1 | 0.1945 (2) | 0.6125 (2) | 0.1898 (2) | 0.0073 (2) |
| Ir2 | 4a | 1 | 0.3463 (1) | 0.1101 (2) | 0.0965 (2) | 0.0075 (2) |
| Nb3 | 4a | 1 | 0.0000 (2) | 0.1115 (4) | 0.0000 (3) | 0.0062 (3) |
| B4 | 4a | 1 | 0.018 (3) | 0.370 (5) | 0.355 (4) | 0.013 (4) |
| B5 | 4a | 1 | 0.198 (3) | 0.195 (5) | 0.354 (4) | 0.011 (4) |

**Table S3.** Anisotropic thermal displacements from NbIr$_2$B$_2$.

| Atom | U11 | U22 | U33 | U23 | U13 | U12 |
|---|---|---|---|---|---|---|
| Ir1 | 0.0061 (3) | 0.0058 (3) | 0.0095 (4) | 0.0028 (7) | 0.0007 (3) | -0.0010 (7) |
| Ir2 | 0.0051 (3) | 0.0056 (3) | 0.0118 (3) | 0.0017 (7) | 0.0018 (2) | -0.0012 (7) |
| Nb3 | 0.0053 (6) | 0.0054 (7) | 0.0078 (7) | 0.0025 (9) | 0.0010 (5) | -0.0006 (9) |
| B4 | 0.019 (10) | 0.011 (10) | 0.011 (10) | 0.006 (8) | 0.005 (7) | 0.007 (8) |
| B5 | 0.008 (8) | 0.011 (9) | 0.011 (10) | 0.004 (7) | -0.002 (7) | 0.006 (7) |

The DFT-relaxed atomic positions for NbIr$_2$B$_2$ and TaIr$_2$B$_2$, see Table S4, do not differ much and are close to the X-ray structure determined refined values.

**Table S4.** Relaxed atomic positions of NbIr$_2$B$_2$ and TaIr$_2$B$_2$

|       | NbIr$_2$B$_2$ | | | TaIr$_2$B$_2$ | | |
|-------|---------|---------|---------|---------|---------|---------|
| **Atom** | *x* | *y* | *z* | *x* | *y* | *z* |
| Ir1   | 0.19449 | 0.61362 | 0.18833 | 0.19558 | 0.61521 | 0.18927 |
| Ir2   | 0.34681 | 0.11110 | 0.09665 | 0.34726 | 0.11356 | 0.09658 |
| Nb/Ta | 0.00119 | 0.11202 | 0.00172 | 0.00087 | 0.11280 | 0.00147 |
| B1    | 0.01434 | 0.37095 | 0.35286 | 0.01348 | 0.36993 | 0.35199 |
| B2    | 0.19995 | 0.19284 | 0.35576 | 0.19963 | 0.19194 | 0.35603 |

**Table S5**. Selected interatomic distances for NbIr$_2$B$_2$.

| Atom1 | Atom2 | Distance (Å) |
|-------|-------|--------------|
| B1    | B2    | 1.69 (4)     |
| Nb    | Ir1   | 2.953 (2) / 2.962 (2) / 2.997 (2) |
| Nb    | Ir2   | 2.761 (2) / 2.815 (2) / 2.819 (3) |
| Ir1   | Ir2   | 2.777 (1) / 2.791 (1) / 2.810 (1) |

Selected interatomic distances in NbIr$_2$B$_2$ are listed in Table S5. Based on the differences in atomic radius, Nb-Ir and Ir-Ir bond lengths are comparable to what are present in previously reported superconducting Nb/TaRh$_2$B$_2$[1]. Turning to the B-B distance, in the rhodium materials it is 1.77(7) Å between two boron atoms in each boron dimer. In our case their separation is 1.69(4) Å which is slightly smaller but still in the standard deviation range. The short B-B distance in the boron dimers may result from the increasing of atomic radii from Rh to Ir, which furthermore compresses the space for boron dimers in between Nb@Ir9 polyhedra. Moreover, considering the quality of our single crystal diffraction results where the highest diffraction peak (2.359 e$^-$/Å$^3$) and deepest diffraction hole (-1.957 e$^-$/Å$^3$) are 1.24 Å from Nb and 0.76 Å from Ir2, respectively, and the Flack parameter is 0.04(2) with small standard deviation, the noncentrosymmetric model in *Cc* space group is highly likely correct. Besides, the powder XRD pattern measured for both compounds are fitted well with the refined crystal structure with no missing peaks which also supports that the model is correct.

**SM-2. Powder X-ray diffraction analysis (pXRD)**

The pXRD patterns with the Rietveld refinement are shown in Figure S1(a) and Figure S1(b) for NbIr$_2$B$_2$ and TaIr$_2$B$_2$, respectively. The pXRD data were analyzed by the Rietveld method with the starting model obtained by a single-crystal refinement. The quantities of impurity phases are as follows: TaB$_2$ (3.5% wt.) and TaIr$_3$ (2.5% wt.) for TaIr$_2$B$_2$ and NbB$_2$ (4.8% wt.) for NbIr$_2$B$_2$. An additional impurity phase is SiO$_2$, which is often present if a boron reach, very hard sample is ground in an agate mortar[2]. The refined lattice and structural parameters (Table S6) are in good agreement with those obtained by the single-crystal X-ray diffraction method.

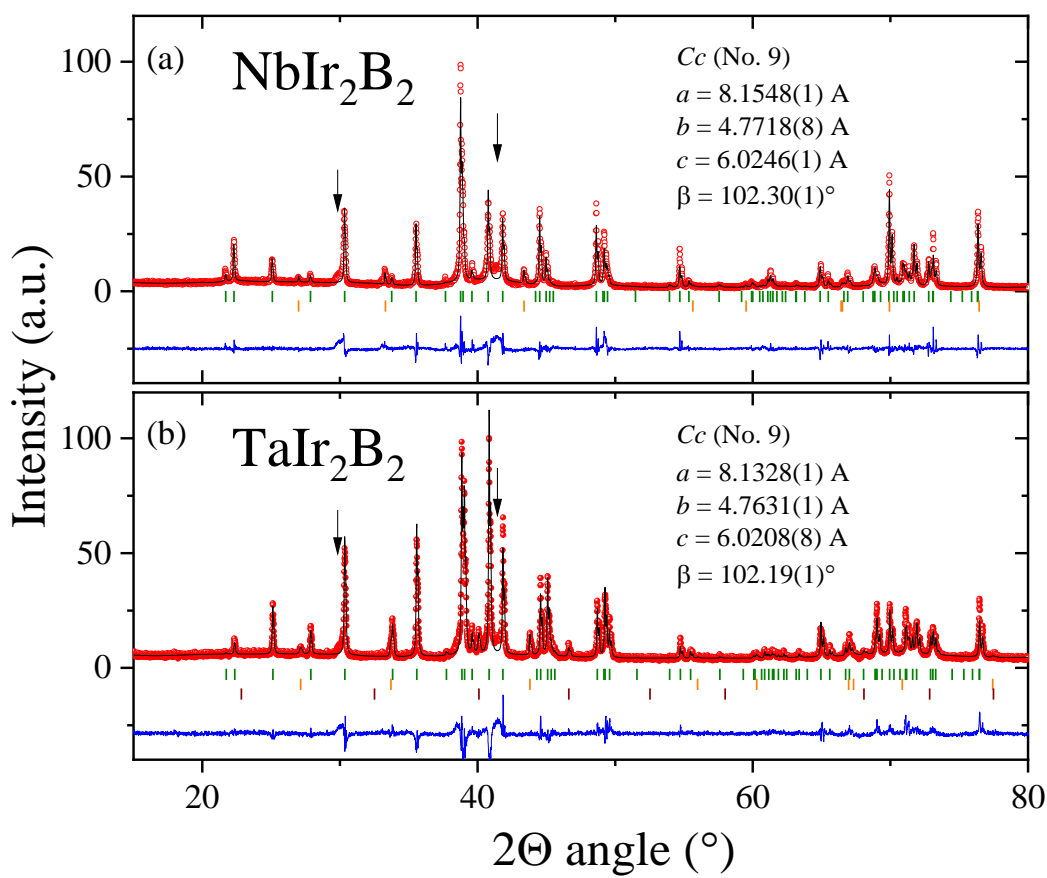

**Figure S1.** Powder x-ray diffraction pattern (pXRD) (red points) together with the Rietveld refinement profile (black solid line) for NbIr$_2$B$_2$ (upper panel) and TaIr$_2$B$_2$ (lower panel). The green, orange and violet vertical bars indicate the expected Bragg peak positions for (Nb/Ta)Ir$_2$B$_2$ (C$c$ (no. 9)), (Nb/Ta)B$_2$ impurity (P6/mmm (no. 191)) and TaIr$_3$ impurity (Pm-3m (no. 221)), respectively. Arrows indicate the broad reflections coming from SiO$_2$ (mortar and pestle). The blue curve is the difference between experimental and model results.

**Table S6.** Refined structural parameters for $NbIr_2B_2$ and $TaIr_2B_2$ obtained from the powder refinements. Background-corrected Rietveld refinement reliability factors: profile residual $R_p$ = 12.4%, weighted profile residual $R_{WP}$ = 16.4%, expected residual $R_{exp}$ = 7.6%, GOF = 2.2 for the Nb analog and $R_p$ = 9.3%, $R_{WP}$ = 12.1%, $R_{exp}$ = 7.0%, GOF = 1.7 for the Ta analog.

|  | $NbIr_2B_2$ | | | $TaIr_2B_2$ | | |
| --- | --- | --- | --- | --- | --- | --- |
| **Atom** | *x* | *y* | *z* | *x* | *y* | *z* |
| Ir1 | 0.22826 | 0.60475 | 0.19022 | 0.19682 | 0.61293 | 0.19139 |
| Ir2 | 0.35654 | 0.10757 | 0.10100 | 0.34676 | 0.11346 | 0.09765 |
| Nb/Ta | 0.00108 | 0.11337 | 0.00142 | 0.00087 | 0.11285 | 0.00147 |
| B1 | 0.01434 | 0.37095 | 0.35286 | 0.01348 | 0.36993 | 0.35199 |
| B2 | 0.19995 | 0.19284 | 0.35576 | 0.19963 | 0.19194 | 0.35603 |

## SM-3. Estimate of the physical properties for NbIr$_2$B$_2$ and TaIr$_2$B$_2$

In order to determine the lower critical field ($\mu_0 H_{c1}^*(0)$), the magnetization was measured as a function of magnetic field at several temperatures below the superconducting transition temperature T$_c$. For each temperature, the experimental data obtained in small magnetic fields were fitted using the proportionality M$_{fit}$ = −pH, appropriate for a full shielding state. Comparing the value of a prefactor p derived from the isotherm taken at T = 1.7 K with the ideal diamagnetism quantified as −1/4$\pi$, the demagnetization factor N = 0.49 for the Nb variant and N = 0.55 for the Ta variant was estimated. Those values are fairly consistent with the expected (theoretical) N$_z$ value derived for a circular cylinder sample with the height to radius ratio of approx. 0.5 (see ref.[3]). The low-field linear fit to the magnetization data (M$_{fit}$) was used to construct the M$_V$ −M$_{fit}$ plot. In the next step, the values of the lower critical field $\mu_0$H$_{c1}^*$(T) were extracted as is shown in Figure S2. Note that the black dashed lines ΔM=M$_V$ −M$_{fit}$= 0.2 emu cm$^{-3}$ (NbIr$_2$B$_2$) and 0.1 emu cm$^{-3}$ (TaIr$_2$B$_2$) were chosen carefully and they are as small as possible (ΔM is less than 2% of the magnetization value M for applied field of H$_{c1}^*$ obtained at T = 1.7 K, for each compound). The resulting values of $\mu_0 H_{c1}^*$ estimated in this manner are depicted in the main panel of Figure 2(c) and Figure 2(d). An additional point for H = 0 is a zero field transition temperature taken from the resistivity measurement.

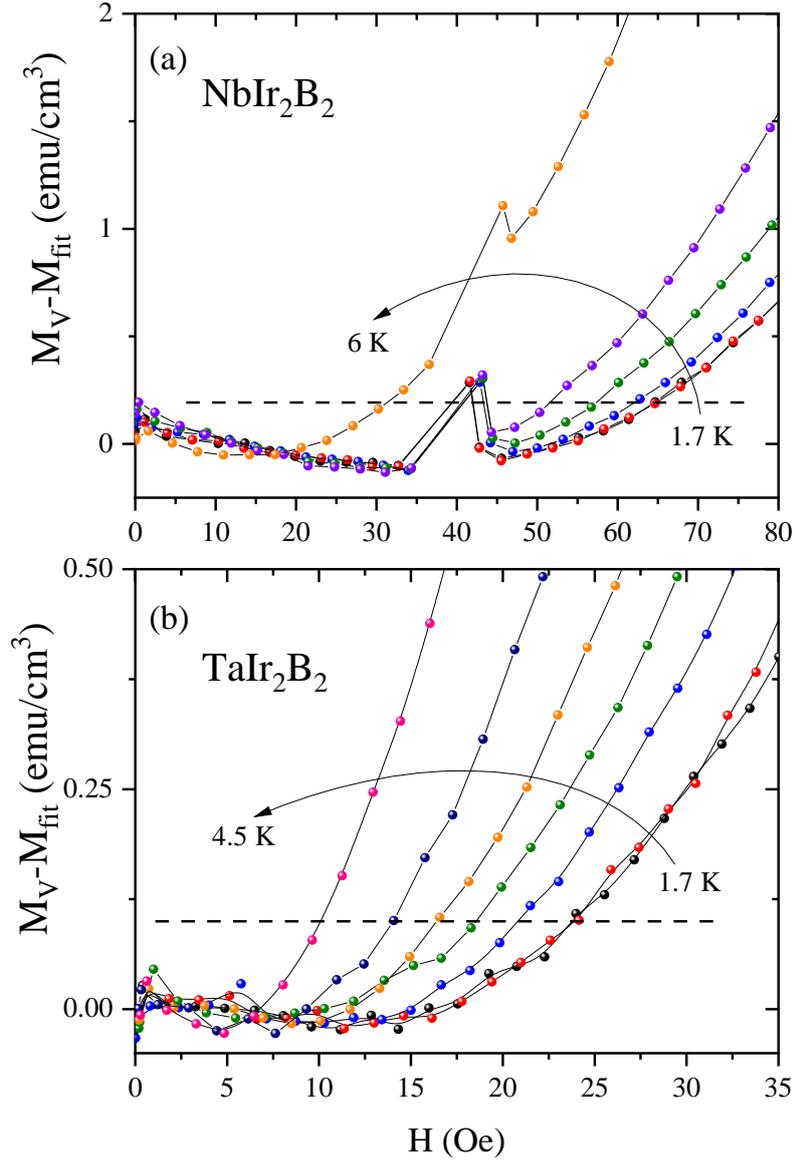

**Figure S2.** The field dependence of the difference between volume magnetization $M_v$ and $M_{fit}$ at various temperatures below $T_c$. The dashed line is a line used for obtaining $\mu_0 H_{c1}^*$ at each temperature. An anomaly seen for H ~40 Oe (part (a)) is possibly caused by a superconducting vortex avalanche effect.

The temperature dependence of the electronic specific heat ($C_{el}$) below $T_c$ for $NbIr_2B_2$ is shown in Figure S3 (a) and (b). $C_{el.}$ was calculated by subtracting the phonon contribution ($C_{ph}$) from the total specific heat $C_p$. The $C_{el}$ was then analyzed by fitting the data with a single gap isotropic *s*-wave model and an isotropic two-gap (*s+s*)-wave model (panel (a)), and power-law model ($C_{el.} \propto T^3$) expected for point nodes (panel (b)). All fits were done below 5K, which is about $0.7T_c$. An *s*-wave single gap BCS model (blue dashed line) gives $2\Delta_0 =$ 2.70(6) meV, with a coefficient of determination $R^2 = 0.9991$. The expected by BCS theory value is $2\Delta_0 = 3.52 k_B T_c = 2.17$ meV. A better fit ($R^2 = 0.99992$) was obtained assuming a multigap ($s+s$) scenario with $2\Delta_{01} = 2.32(5)$ meV and $2\Delta_{02} = 9.1(12)$ meV, represented by a red line. A solid and dash line in an inset represent difference between experiment and a single and double $s$-wave gap model, respectively. For a gap with nodes, theory predicts power-law dependence, with n = 2 or 3 for line or point nodes, respectively. A green line in a panel (b) is a $C_{el.} \propto T^3$ fit, and its low quality does not support a point nodes scenario for NbIr$_2$B$_2$. A better fit ($R^2 = 0.9988$), but still worse than obtained for the gap models, was obtained for $C_{el.} \propto T^n$ and a refined n value is 3.65(2), which is close to n = 3.3 reported for W$_3$Al$_2$C[4]. The difference between experiment and power-law models are shown in an inset. Note that the scale here is twice larger comparing to an inset in an upper panel.

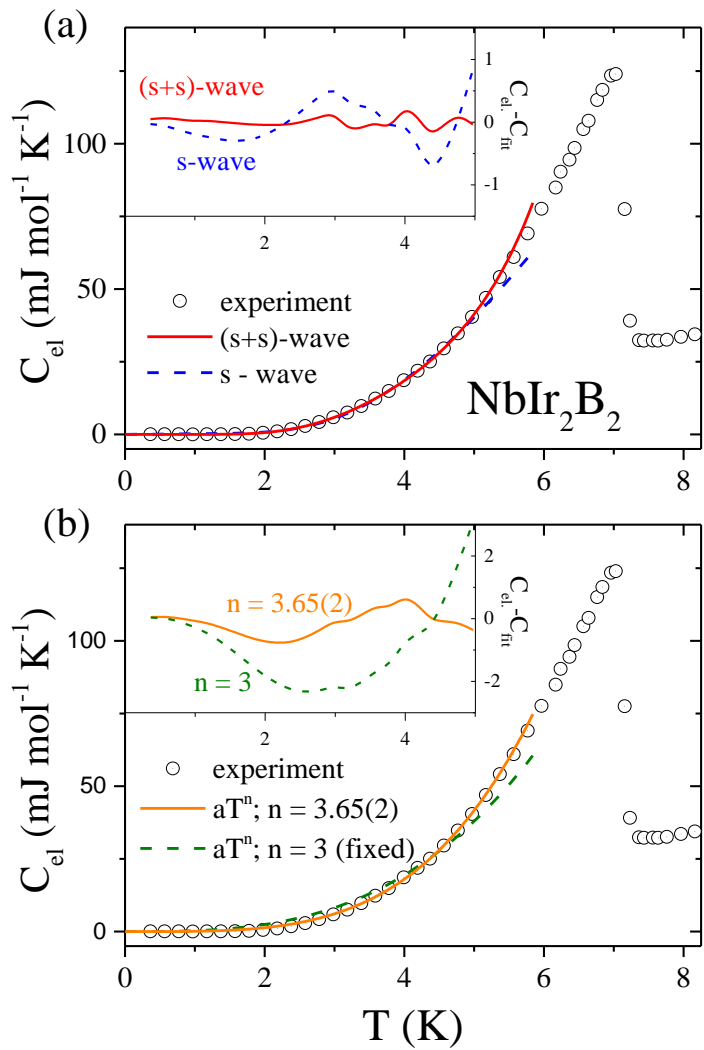

**Figure S3.** Temperature-dependent electronic specific heat $C_{el.}$ for NbIr$_2$B$_2$. (a) a fit of a single gap isotropic $s$-wave model (blue dashed line) and an isotropic two-gap ($s+s$)-wave model (red solid line) to the data. (b) a fit of a power-law model ($C_{el.} \propto T^3$) to the data.

The last experimental technique used for characterization of the new superconductors was temperature dependent resistivity. NbIr$_2$B$_2$ behaves like a poor metal, with a shallow negative gradient for the resistivity upon cooling from room temperature. The residual resistivity ratio (RRR), $\rho_{300}/\rho_{10}$ = 1.3, is small, which can be attributed to the polycrystalline nature of the sample contained grain boundaries and macroscopic defects. In the case of Ta variant, one observes an increase in $\rho(T)$ as the temperature was decreased. Comparing $\rho$(300 K) and $\rho$(10 K), resistivity increases about 50%. A non-metallic behavior is likely caused by disorder scattering and the charge carrier localization effect. At low temperatures the electrical resistivity drops sharply to zero at $T_c$ = 7.24 K for NbIr$_2$B$_2$ and at $T_c$ = 5.38 K for TaIr$_2$B$_2$. Superconducting critical temperature $T_c$ is defined by the temperature of the 50% drop of the $\rho(T)$ data in zero magnetic field. The slightly higher superconducting temperature value obtained in the resistivity measurement is likely due to the influence of surface superconductivity emerging in each cross-sectional area of the sample. The effect of applying a magnetic field on $T_c$ is shown in the inset of Figure 3(d) for Nb analog and Figure 3(e) for Ta analog. As expected, the transition becomes broader, and $T_c$ shifts to lower temperature as the applied field was increased. It should be noted that a transition to a zero-resistance state was obtained even at 9 T and above 3 K for NbIr$_2$B$_2$ or 2 K for TaIr$_2$B$_2$, indicating a large upper critical field.

To estimate the upper critical field at 0 K, $\mu_0H_{c2}(0)$, we fit the data shown in Figure 3(f) to the Ginzburg-Landau (GL) expression

$$\mu_0 H_{c2}(T) = \mu_0 H_{c2}(0) \frac{(1-t^2)}{(1+t^2)} \qquad (1)$$

where t = $T/T_c$ and $T_c$ is the transition temperature at zero magnetic field. The fit of the resistivity data is represented by a solid line and the obtained values of $\mu_0H_{c2}(0)$ are: 16.3(2) T and 14.7(1) T for NbIr$_2$B$_2$ and TaIr$_2$B$_2$, respectively. For the thermodynamic data the values of the upper critical fields are $\mu_0H_{c2}(0)$ = 15.8(1) T for Nb variant and $\mu_0H_{c2}(0)$ = 16.5(2) T for Ta variant (the dashed line). Furthermore, a conservative evaluation of the upper critical fields at 0 K was performed using the Werthamer-Helfand-Hohenberg (WHH) approximation in the dirty limit[5,6]:

$$\mu_0 H_{c2}(0) = -A T_c \left.\frac{d\mu_0 H_{c2}}{dT}\right|_{T=T_c} \quad (2)$$

where A is the purity factor given by 0.693 for the dirty limit. Table S7 presents the expected $\mu_0 H_{c2}(0)$ value obtained from WHH and GL models.

**Table S7.** The upper critical field $\mu_0 H_{c2}(0)$ estimated from WHH and GL models.

| $\mu_0 H_{c2}(0)$ | Unit | NbIr$_2$B$_2$ | TaIr$_2$B$_2$ |
|---|---|---|---|
| WHH $_{\text{resistivity}}$ | T | 13.3(2) | 11.6(1) |
| GL $_{\text{resistivity}}$ | T | 16.3(2) | 14.7(1) |
| WHH $_{\text{heat capacity}}$ | T | 14.2(2) | 12.8(1) |
| GL $_{\text{heat capacity}}$ | T | 15.8(1) | 16.5(2) |
| $\mu_0 H_{c2}^p(0)$ | T | 13.3(1) | 9.5(1) |

## SM-4. Electronic structure calculations

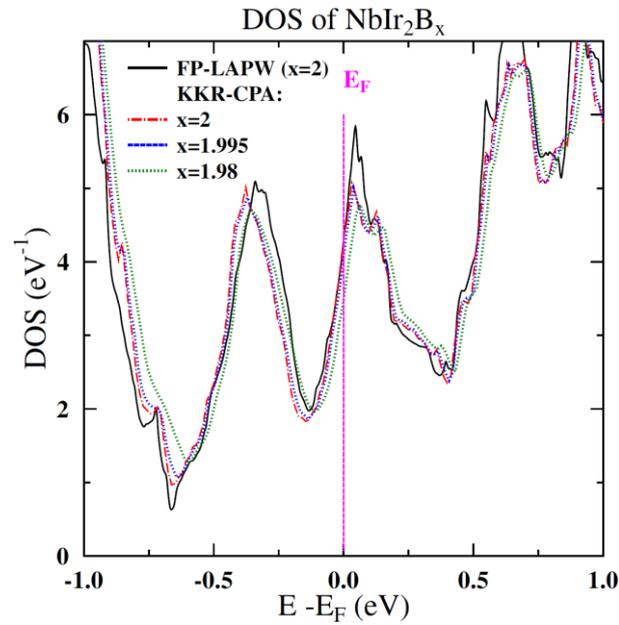

**Figure S4**. Evolution of DOS near the Fermi level in NbIr$_2$B$_x$ as a function of boron concentration, obtained using the KKR-CPA method. For the stoichiometric composition (x = 2) KKR-CPA DOS very well agrees with the FP-LAPW result.

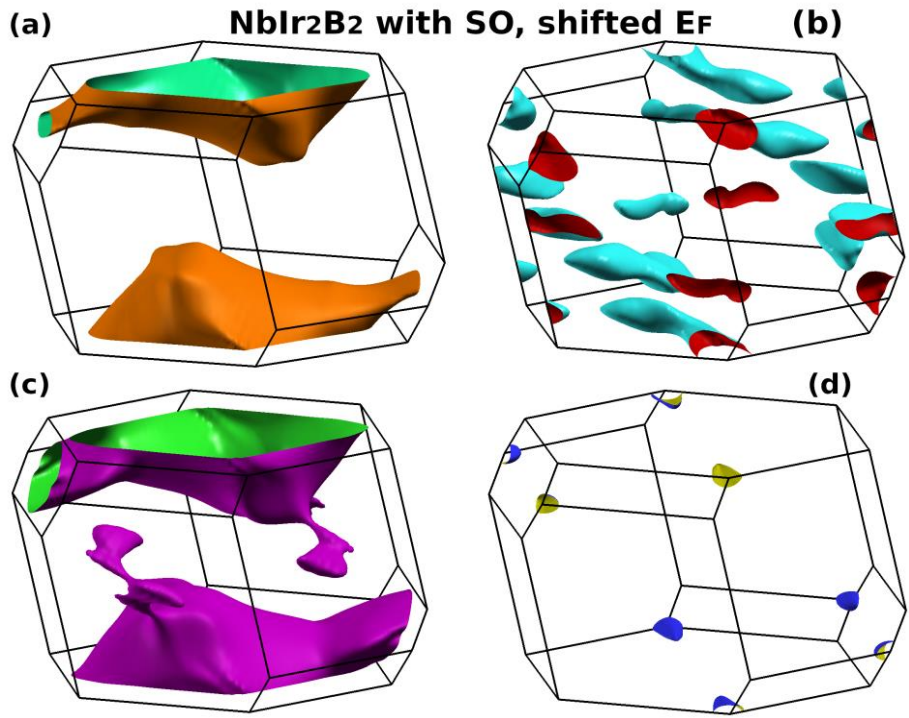

**Figure S5.** Fermi surface of NbIr$_2$B$_2$ calculated with spin-orbit coupling, for the shifted Fermi energy, so as the experimental and renormalized calculated Sommerfeld parameters match. E$_F$ shift, comparing to Fig. 5 is equal to -84 meV. The contribution from the third and fourth FS sheet (panels b,d) is considerably reduced, as compared to Fig. 5 (d,f).

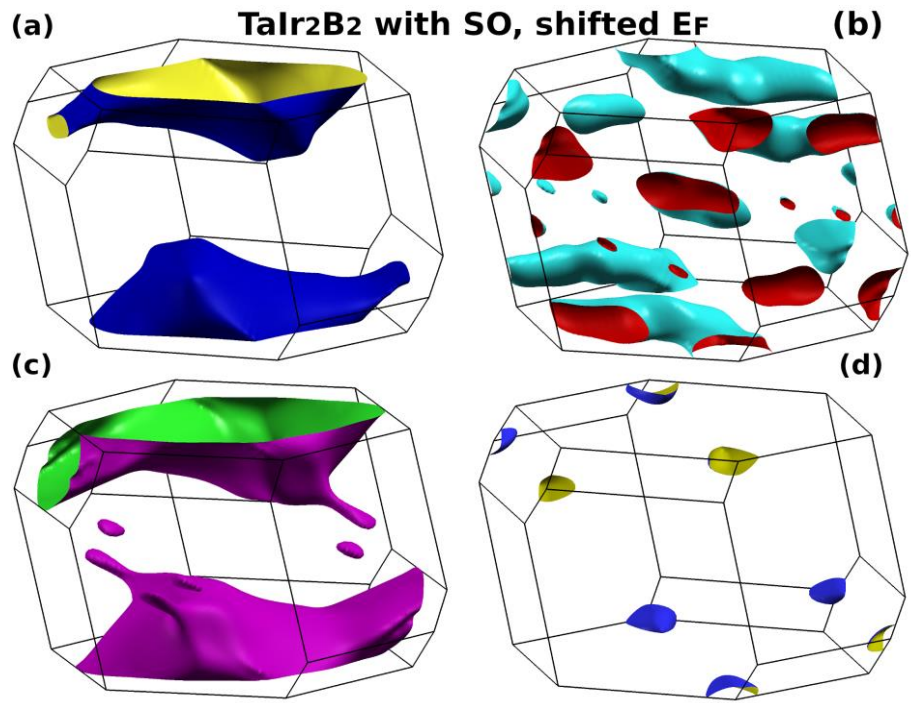

**Figure S6.** Fermi surface of TaIr$_2$B$_2$ calculated with spin-orbit coupling, for the shifted Fermi energy, so as the experimental and renormalized calculated Sommerfeld parameters match. E$_F$ shift, comparing to Fig. 5 is equal to -73 meV. The contribution from the third and fourth FS sheet (panels b,d) is considerably reduced, as compared to Fig. 5 (n,p).

**References**

[1] E. M. Carnicom, W. Xie, T. Klimczuk, J. Lin, K. Górnicka, Z. Sobczak, N. P. Ong, R. J. Cava, *Science Advances* **2018**, *4*, eaar7969.
[2] E. M. Carnicom, J. Strychalska-Nowak, P. Wiśniewski, D. Kaczorowski, W. Xie, T. Klimczuk, R. J. Cava, *Superconductor Science and Technology* **2018**, *31*, 115005.
[3] M. Sato, Y. Ishii, *Journal of Applied Physics* **1989**, *66*, 983.
[4] T. P. Ying, Y. P. Qi, H. Hosono, *Physical Review B* **2019**, *100*, 094522.
[5] N. R. Werthamer, E. Helfand, P. C. Hohenberg, *Phys. Rev.* **1966**, *147*, 295.
[6] E. Helfand, N. R. Werthamer, *Phys. Rev.* **1966**, *147*, 288.